\documentclass[aps,prx,twocolumn,groupedaddress]{revtex4-1}

\usepackage{color}

\usepackage[utf8]{inputenc}
\usepackage[english]{babel}
\usepackage{amsmath,graphicx,enumerate}

\begin{document}

\title{How a local active force modifies the structural properties of polymers} 

\author{L. Natali$^{1}$}

\author{L. Caprini$^{1}$}

\author{F. Cecconi$^{3}$} 
\affiliation{$1$, Dipartimento di Fisica, Universit\`{a} ``Sapienza'', Piazzale A. Moro 5, I00185 Rome Italy.\\
$^2$ Gran Sasso Science Institute (GSSI), Via. F. Crispi 7, 67100 L'Aquila, Italy. E-mail: lorenzo.caprini@gssi.it\\ 
$^3$, Istituto dei Sistemi Complessi-CNR, Via Taurini 19, I-00185 Rome Italy. }

\date{\today}

\begin{abstract}
We study the dynamics of a polymer, described as a variant of a Rouse chain, driven by an active terminal monomer (head).
The local active force induces a transition from a globule-like to an elongated state, as revealed by the study of the end-to-end distance, whose variance is analytically predicted under suitable approximations.
The change in the relaxation times of the Rouse-modes produced by the local
self-propulsion is consistent with the transition from globule to elongated 
conformations. Moreover also the bond-bond spatial correlation for the 
chain head results to be affected and 
a gradient of over-stretched bonds along the chain is observed.
We compare our numerical results both with the phenomenological stiff-polymer theory and several analytical predictions in the Rouse-chain approximation.
\end{abstract}

\maketitle

\section{Introduction}
Active matter includes a large class of physical and biological entities
ranging from microscopic to macroscopic length scales.
Active systems usually convert fuel energy from the environment into
directed motion using chemical reactions~\cite{bechinger2016active} or
propelling in a fluid through flagella, cilia or more complex
mechanisms~\cite{elgeti2015physics}.
Although numerous studies have focused on spherical rigid or rod-like microswimmers, many active systems, appears in polymeric and filamentous structures which 
often undergo stretching or deformations. 
In particular, the cell cytoskeleton contains many active filaments, like actins and microtubules~\cite{ridley2003cell,ganguly2012cytoplasmic}. 
Recently, microtubules have been studied in vitro and transported along a 
glass substrate by ATP fueled motor proteins~\cite{weber2015random}.

Out of the biological realm, it is even possible to realize synthetic 
active colloidal polymers~\cite{lowen2018active}, such as chains of 
colloids uniformly coated with catalytic particles~\cite{biswas2017linking} 
becoming active when immersed in a solution of $H_2 O_2$.
The ``activation'' of Janus particles can be even controlled by external 
fields, and, in particular, 
the spontaneous formation of chains of particles has been experimentally 
observed upon tuning the frequencies of an AC electric field~\cite{di2016active,yan2016reconfiguring,nishiguchi2018flagellar}. 

The study of active or activated flexible and semi-flexible polymers 
has received much attention in the last years, for both biological interest 
and possible applications towards the design of new materials with peculiar 
properties.
Several authors, via computer simulations, addressed not only the 
``activation'' of polymers~\cite{Mackintosh1997,Eisenstecken2016,Kaiser2015}  
but also studied the behavior of polymer chains immersed into an active bath~\cite{Shin2015,Chaki2019,Kaisera2014,Nikola2016,Harder2014, samanta2016chain}. 
Sometimes, the activation is modeled by imposing a self-propulsion force 
tangential to the filament~\cite{Ghosh2014,Isele2015,Isele2016,Laskar2017,Chelakkot2012}, while the common approach for the effect of an active bath 
amounts to considering the monomer under independent active 
forces~\cite{Kaisera2014, Kaiser2015, Eisenstecken2016, liu2019configuration2}. 
In any case, the interplay between active forces and extended flexible 
structures gives rise to a rich phenomenology also including collective 
behaviors~\cite{Rodriguez2003, DeCamp2015, Sumino2012, Schaller2011, Schaller2013, prathyusha2018dynamically, vliegenthart2019filamentous}.
For instance, a freely moving active filament takes on peculiar dynamical
conformations performing: rotational, straight translational~\cite{Jiang2014,Sarkar2016}, snake-like~\cite{Isele2015,Isele2016} and even helical motion~\cite{anand2018structure}. 
In addition, in the limit of large active forces, both semi-flexible and flexible polymers swell~\cite{Eisenstecken2016,Kaiser2015} 
while clamped filaments exhibit beating and rotational motion under tangential 
forces~\cite{Laskar2017,Chelakkot2014, anand2019beating}. 
In some cases, compactification and shrinkage of structures occur at very 
strong active forces~\cite{bianco2018globulelike,duman2018collective}, 
and at high densities swirls and spirals are induced by the increase of the 
active force~\cite{prathyusha2018dynamically}. 
Some authors implemented the activity, generated by the release of energy due to ATP hydrolysis, as a temperature increase observing 
phase-separations in binary mixture of passive-active polymers \cite{smrek2017small, smrek2018interfacial}.

The studies mentioned above deal with global activated polymers. 
However, there are biological examples in which the self-propulsion is 
generated only in local regions of the active particle. 
For instance, some elongated bacteria move thanks to cilia 
attached to specific regions of their body, analogously, spermatozoa swim 
due to single flagellum protruding from the body. 
Other examples come from the action of RNA polymerase on DNA or kinesin on 
microtubules, which are generally described as molecular motors on 
polymer substrates~\cite{foglino2019non}. Thus, for long and flexible 
swimmers, we need to go beyond the collective activation or center of mass
description, because the interplay between deformability and self-propulsion 
induces a richer phenomenology.

Some authors considered  polymers with a catalytic terminal (head)\cite{Tao2009,
Sarkar2015}, where the chemical reactions occurring at the head produce 
a local self-propulsion which increases the effective diffusivity of the chain.
 
Motivated by these works, we study the activation of the 
terminal monomer of a polymer, modeling the self-propulsion in the framework of 
non-equilibrium stochastic processes.
In particular, we adopt a well-established model, known as Active Ornstein-Uhlenbeck (AOUP) model \cite{marconi2016velocity, fodor2018statistical, caprini2018linear, wittmann2019pressure, berthier2019glassy, caprini2019} to describe the self-propulsion at a coarse-grained level, neglecting the microscopic details of the active force. 
We describe the behavior of the polymer in the absence of any 
confinement or external potential to determine how a local active force 
affects the chain conformations.  

This paper is organized as follows.
In Sec.~\ref{sec:model}, we introduce the model describing the polymer in the 
presence of a local active force, activated only on the last monomer, while, in Sec.~\ref{sec:freepolymer}, we present 
numerical results in the absence of any external mechanism or molecular motor.
We focus, on one hand, on the study of the end-to-end distance and, on
the other hand, on the structural microscopic properties of the polymer, unveiling the effect of the local active force.
Finally, we summarize the results and discuss some future perspectives 
in the conclusive Section.

\section{A free polymer with an active head
\label{sec:model}}
To model a polymeric structure, such as proteins or biological 
filaments, we employ a variant of the Rouse-chain~\cite{Rouse}, 
assuming that each monomer has the same structure and composition. 
Each monomer is only connected to the nearest neighbors by springs of strength $k$ and rest length $\sigma>0$. 
Since we neglect the steric interactions among non-consecutive beads 
the polymer is fully described by the simple potential:
\begin{equation}
\label{eq:PotentialPolymer}
U(\mathbf{r}_{1},\ldots,\mathbf{r}_{N})= \frac{k}{2}\sum_{n=1}^{N-1}
	(| \mathbf{r}_{n+1}- \mathbf{r}_{n} | - \sigma )^{2} \, \,
\end{equation}
and its dynamics is ruled by $N$ coupled Langevin equations for the positions and velocities of 
each monomer, $\textbf{r}_n$, and $\textbf{v}_n$, respectively:
\begin{subequations}
\label{eq:polymerdynamics}
\begin{align}
\dot{\mathbf{x}}_{n} &= \mathbf{v}_{n} \\
\dot{\mathbf{v}}_{n} &= -\frac{\mathbf{v}_{n}}{\tau_0} - 
	\frac{\partial U}{\partial \mathbf{r}_{n}} + \frac{\sqrt{2D_t}}{\tau_0} \boldsymbol{\xi}_{n} + \delta_{n,N}{\mathbf f}_a\,,
\end{align}
\end{subequations}
where $\tau_0$ is the inertial relaxation time of each monomer. 
$\boldsymbol{\xi}_n$ is a white noise vector whose uncorrelated components have zero averages and unit variances, 
while $D_t$ is the diffusion coefficient due to the solvent.
The last term, $\delta_{n,N} \textbf{f}_a$, represents the active force, 
due, for instance, to ATP-hydrolysis or other chemical reactions occurring at a 
catalytic site. Since we assume that the reaction takes place on 
one terminal only, we denote such a monomer as ``catalytic head''.

To describe the fluctuations of the active force, 
we employ the Ornstein-Uhlenbeck process (AOUP model)
\begin{equation}
\label{eq:activeforcedynamics}
\tau\dot{\mathbf{f}}_a=-\mathbf{f}_a +\sqrt{2D_a} \boldsymbol{\eta} 	\,,
\end{equation}
being $\boldsymbol{\eta}$ a white noise vector with zero averages and unit 
variances. 
The two-time activity-activity correlation of $\mathbf{f}_a$ decays exponentially, with a correlation time, $\tau$,
that roughly determines the time-window after which the active force completely resets its value.
We remark that the AOUP model constitutes a simplification of the Active Brownian Particle model 
(ABP)~\cite{fily2012athermal, marchetti2013hydrodynamics, basu2019long1, siebert2017phase} which is known to explain the well-known 
phenomenology of spherical self-propelled particles~\cite{fodor2018statistical,fodor2016far,caprini2018active, caprini2019entropy, PhysRevLett.122.258001}. 
The connection between AOUP and ABP has been shown by several authors~\cite{das2018confined,caprini2019comparative1}.
In Eq.\eqref{eq:activeforcedynamics}, the parameter $\sqrt{D_a/\tau}$ has a particular relevance because it sets the 
strength of the self-propulsion. In other terms, 
a single particle performs a persistent motion in the 
direction of $\mathbf{f}_a$ for a time $t<\tau$, while for $t> \tau$ a 
diffusive-like behavior of a passive Brownian particle with a certain effective temperature is recovered.  
When $\tau$ is the smallest time scale in the system, the active force can be simply recast into a Brownian motion 
$\textbf{f}_a\approx\sqrt{2D_a}\boldsymbol{\eta}$, being $\textbf{f}_a$ the faster degree of freedom whose 
time-derivative could be set to zero.
We expect that in such a case the head does not display any persistence and the role of the active force leads just 
to the increase of the effective diffusion.
In addition, in order to enhance the effect of the self-propulsion, we focus on a regime where the velocity of the monomers 
relaxes faster than $\textbf{f}_a$, meaning that the inertial time $\tau_0$ is smaller than $\tau$.

The dynamics of the  polymer  center of mass, 
$\mathbf{r}_c=\sum_n \mathbf{r}_n/N$ and $\textbf{v}_c=\sum_n \textbf{v}_n/N$, 
could be simply obtained by summing up Eqs.\eqref{eq:polymerdynamics} for all the monomers:
\begin{subequations}
\label{eq:centerofmass_maintext}
\begin{align}
\label{eq:Xc}
\dot{\mathbf{r}}_c &= \textbf{v}_c \\
\label{eq:Vc}
	\dot{\mathbf v}_c &= - \frac{\textbf{v}_c}{\tau_0} + 
	\dfrac{\mathbf f_a}{N} + \frac{\sqrt{2 D_t}}{\tau_0\sqrt{N}}\,
	\boldsymbol{\xi} \,,
\end{align}
\end{subequations}
being $\boldsymbol{\xi}$ a new white noise vector whose uncorrelated components have unit variance and zero average.
The center of mass behaves as a free active particle, where the amplitude of the effective bath scales as $1/\sqrt{N}$ and 
the active force is decreased by a factor $N$. 
The linearity of Eq.\eqref{eq:Xc} and Eq.\eqref{eq:Vc} allows us to find 
the joint probability distribution function of the velocity and the active force of the center of mass: 
\begin{equation}
p\left(\textbf{v}_c, \textbf{f}_a \right) 
\propto G(\mathbf{f}_a)\;\exp{\left[-\frac{\beta_{\mathrm{eff}}}{2}\left( \mathbf{v}_c - \mathcal{C}\,\mathbf{f}_a \right)^2\right]}\,.
\end{equation}
where $G(\mathbf{f}_a)$ is a Gaussian function centered in zero and the coefficients $\beta_{eff},\mathcal{C}$ both depend on $\tau$, $\tau_0$, 
$D_a$ and $D_t$, see Appendix~\ref{app:A}.
This analysis shows that, given an $\mathbf{f}_a$, $\mathbf{v}_c$ assumes a typical average value  
$\langle \mathbf{v}_c \rangle\propto\mathbf{f}_a$, thus in a time window smaller than $\tau$, the polymer is driven by the active force 
whose value is extracted by a Gaussian distribution.

The possibility of the active force to drag the polymer can be easily 
estimated by comparing the rescaled variance of $\textbf{f}_a$ with the 
variance of the thermal bath, in Eq.\eqref{eq:Vc}.  
The active force on the head to be effective needs to overwhelm the 
thermal agitation of the passive polymer, thus 
\begin{equation}
\frac{D_a}{\tau\, N^2} \gg \frac{D_t}{N\, \tau_0^2}\,. 
\label{eq:condition} 
\end{equation}
This condition follows from the exact formula of the mean square displacement,
$\mathrm{MSD}(t)$, of the polymer center of mass derived in 
Appendix~\ref{app:A},
\begin{equation}
{\mathrm{MSD}}(t) =  6 \bigg(\frac{D_t}{N} +  
\frac{D_a\tau^2_0}{N^2}\bigg)\;t + 
6 \frac{D_a \tau^2_0}{N^2} \tau \big(e^{-t/\tau} - 1\big) \,.
\label{eq:MSDmaintext}
\end{equation}
The linear term dominates in the   
long-time limit, therefore the active force is able to affect the diffusive 
dynamics of the center of mass only if $D_a/N \simeq D_t/\tau_0^2$. 
Instead, for small times, the active force produces a ballistic
contribution in the ${\mathrm{MSD}}(t)$ as it can be deduced 
by expanding the exponential in power of $t/\tau$ up to the second order. 
Even for small times, this term is relevant with respect to the diffusive one 
if the condition $D_a/\tau/N^2 \gg D_t/N/\tau_0^2$ holds.
Throughout the rest of the paper, we assume that both conditions 
are satisfied: the first increasing the center of mass diffusivity and the 
second leading to a ballistic time regime.
\begin{figure}[!t]
\includegraphics[clip=true,width=\columnwidth,keepaspectratio]
{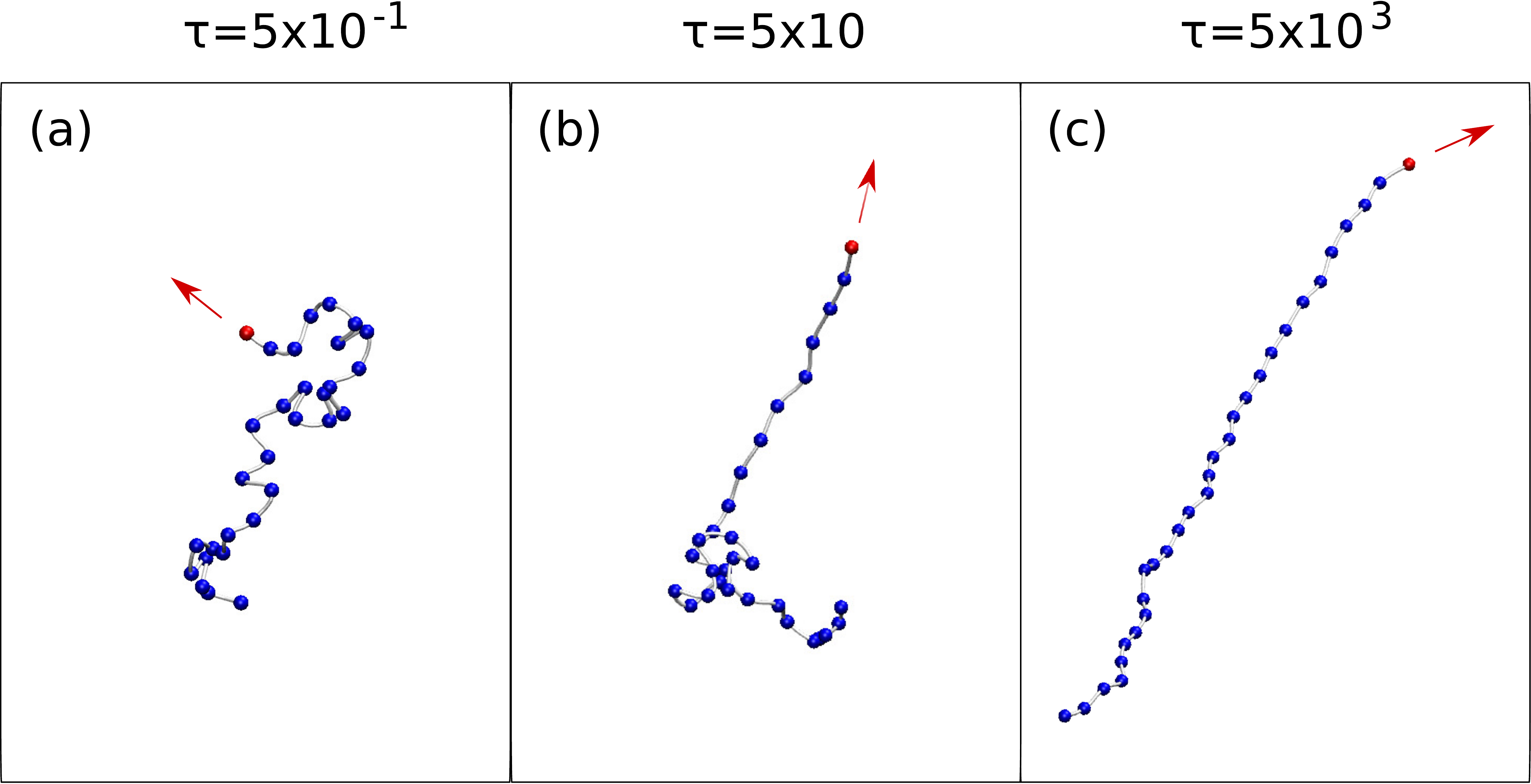}
\caption{Conformations for three different values of $\tau=5\times10^{-1},\, 5\times10,\, 5\times10^4$, 
from the left to the right. Each monomer is represented as a sphere centered 
in $\mathbf{r}_i$ and the grey links among beads are just a guide for the eyes. 
The head of the polymer is the red monomer, whose active force is drawn as a 
red vector, while the passive monomers are colored blue.
The remaining parameters are $D_a/\tau=10^2$, $\sigma=5$, $k=10$, $T=1$ and $\tau_0=0.1$.
Simulations are obtained using a time-step $\sim 10^{-4}$ and each configuration is evolved, at least, for a final time $\sim10^2\tau$.}
\label{fig:snapshot}
\end{figure}

\section{Effect of a local active force on a free polymer
\label{sec:freepolymer}}
Understanding the polymer dynamics beyond Eqs.(\ref{eq:Xc},\ref{eq:Vc})  
requires numerical integration of Eqs.~\eqref{eq:polymerdynamics}. 
To simulate the dynamics, we employ a stochastic Leapfrog 
algorithm~\cite{Burrage2007} and focus on some typical observables able 
to unveil the interplay between the active force and the polymer 
deformations.
Setting a large value of $D_a/\tau$ to satisfy the 
condition~\eqref{eq:condition}, we explore a range of small and large 
$\tau$ compared to the relaxation times of the Rouse modes
of the passive chain \cite{Rouse} 
\begin{equation}
	\tau_p = \dfrac{1}{4 k \tau_0 
	\sin^2 \bigg(\frac{p\pi}{2N}\bigg)}\,.
\label{eq:taup}
\end{equation}

In Fig.\ref{fig:snapshot}, we plot three snapshots of the
conformations of a polymer with $N=30$ monomers at different values of $\tau$.
Each panel corresponds to $D_a/\tau=10^2$, at which the polymer center of mass is transported by the active force.
In panel (a), when $\tau$ is small, the polymer behaves as a passive system, being $\mathbf{f}_a \approx \sqrt{2D_a}\boldsymbol{\eta}$.
In this case, indeed, the active force only superimposes an additional
Brownian motion to the dynamics of the head.
Therefore, the polymer swells a bit, however maintaining the well-known 
coiled structure of a passive Rouse-chain polymer~\cite{Rouse}.
Increasing $\tau$, the head starts to pull some of the 
monomers which protrude from the coil along the direction pointed by 
the active force, in such a way that an elongated portion of the chain 
coexists with the remaining coiled portion.
In this case, the head has enough time to carry the center of mass and 
the polymer displays a persistence dynamics in one direction which is 
slowed down by the ''passive'' globule.
\begin{figure*}[!t]
\includegraphics[clip=true,width=0.75\textwidth]
{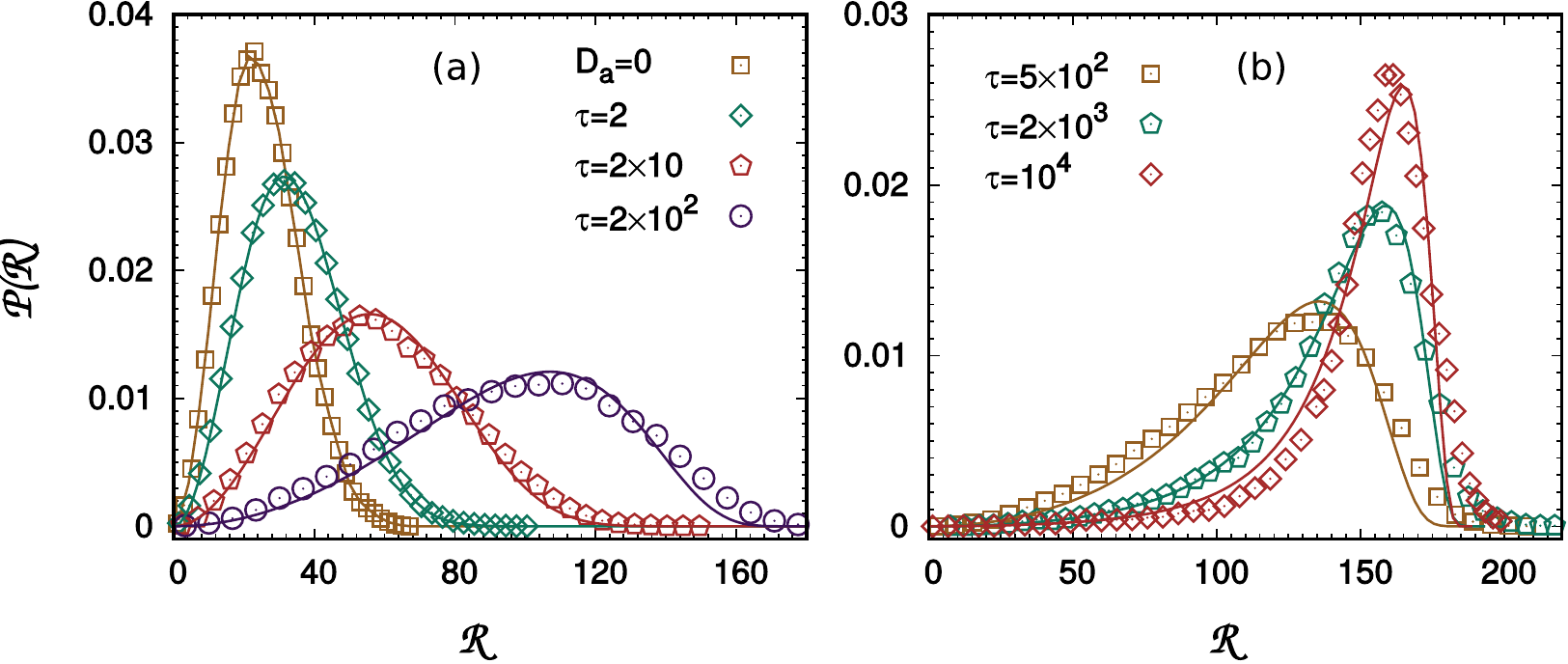}
\caption{Probability distribution of the end-to-end distance, 
$\mathcal{P}(\mathcal{R})$, for different values of $\tau$. Points are obtained via numerical simulations while continuous lines from Eq.\eqref{eq:nearequilibrium_endtoend}.
Data are separated into two panels for presentation reasons: panel (a) 
refers to small values of $\tau$ while panel (b) refers to large values of $\tau$. 
The remaining parameters are $D_a/\tau=10^2$, $\sigma=5$, $k=10$, $D_t=0.1$ and $\tau_0=0.1$.
Simulations are obtained using a time-step $\sim 10^{-4}$ and each configuration is evolved, at least, for a final time $\sim10^2\tau$.
\label{fig:multiplot2}
}
\end{figure*}
A further increase of $\tau$, panel (c), leads the polymer to be fully 
elongated in a rod-like conformation which is carried by the active head. 
The dynamics of the polymer reveals a time-persistence along the random direction pointed by the active force. 
When the head changes direction (roughly after time $\sim \tau$) the rest 
of the monomers follows the head, turning with a typical time delay 
depending on the distance from the head.

To characterize the ``global'' effect of a local active force on the polymer dynamics, we will monitor the distribution of the end-to-end distance focusing on its moments. 
Then, we will also explore the ``local'' effect by quantifying the degree 
of the stretching produced along the chain by the active head. 
Finally, we will focus on the velocity of the head, revealing a bump 
in its variance with the increase of the persistence time.

\subsection{Macroscopic properties of the head-active polymer}
To quantify the degree of elongation taken by the polymer, 
we consider the end-to-end distance
\begin{equation}
\mathcal{R}=|\mathbf{r}_N-\mathbf{r}_1| \,,
\end{equation}
and its distribution
$$
P(\mathcal{R}) = \langle 
\delta(\mathcal{R} - |\mathbf{r}_N-\mathbf{r}_1|) \rangle \,,
$$
at different values of $\tau$ and for fixed $D_a/\tau=10^2$. 
The end-to-end distance, like the gyration radius, is an important observable in polymer physics, providing an estimate of polymer sizes.
Its distribution is experimentally accessible by FRET spectroscopy~\cite{Soranno2009,OBrien_FRET}, moreover, the end-to-end distance is involved in the 
mechanism of polymer looping and cyclization~\cite{Cyclization}. 

Panel~(a) of Fig.\ref{fig:multiplot2} plots $\mathcal{P}(\mathcal{R})$ from numerical 
simulations at fixed $D_a/\tau=10^2$ but
different values of $\tau$ in the interval $0 \div 2\times 10^2$, 
corresponding to the range of small persistence.  
For comparison, we show also the passive case ($D_a=0$) that, 
in the limit $N\gg 1$, is simply described by the well known 
Gaussian-like shape
\begin{equation}
\mathcal{P}(\mathcal{R}) \propto \mathcal{R}^{2}
\exp{\left(-\frac{3}{2}\frac{\mathcal{R}^{2}}{\langle\mathcal{R}^2\rangle}
\right)} \,,
\label{eq:nearequilibrium_endtoend}
\end{equation} 
where $\langle \mathcal{R}^2\rangle$ is the second moment of the 
distribution. Expression~\eqref{eq:nearequilibrium_endtoend} follows as 
a simple consequence of the central limit theorem.
In the low-activity regime, the distribution remains Gaussian-like but 
with a variance increasing with $\tau$, indicating that, in this regime, 
the active force is only able to induce a renormalization of the 
diffusion coefficient.
Panel~(b) of Fig.~\ref{fig:multiplot2} reports the 
$\mathcal{P}(\mathcal{R})$ in the large persistence regime with $\tau$ ranging within $5\times 10^2\div 10^4$. 
Starting from $\tau \approx 5 \times 10^2$ strong non-Gaussian effects appear 
in the shape of $\mathcal{P}(\mathcal{R})$ and 
Eq.\eqref{eq:nearequilibrium_endtoend} is no longer a reasonable approximation.
Specifically, the peak shifts towards larger values of $\mathcal{R}$,  
the longest tail occurs at small $\mathcal{R}$ and accordingly 
the skewness of the distribution changes sign. 
A further increase of $\tau$ narrows the distribution and makes it more 
peaked around $\mathcal{R} = (N-1)\sigma$, corresponding to the end-to-end 
distance of the entirely elongated chain.  
Interestingly, the main peak for $\tau > 10^3$ occurs for 
$\mathcal{R} > (N-1)\sigma$ indicating that the chain is not only 
elongated but also over-stretched. 
In this stretched regime, $\mathcal{P}(\mathcal{R})$ weakly depends on $\tau$ and its increase produces very small changes in the distribution 
until a delta-like shape is achieved at very high $\tau$.

The persistence of the active force confers to the chain a certain spatial
persistence starting from the active head.
This suggests fitting the numerical distributions via the formula
\begin{equation}
	\mathcal{P}(\mathcal{R}) 
	\propto \dfrac{4\pi \mathcal{R}^2}{L^2 - \mathcal{R}^2}
	\exp \bigg\{-\dfrac{9 L^3}{\ell_p(L^2 - \mathcal{R}^2)}\bigg\} \,,
	\label{eq:stiff}
\end{equation}
that was derived for stiff polymers \cite{Thirumalai}.
Where $\ell_p$ is the effective persistence length of the chain and 
$L$ is the maximal contour length that includes possible overstretching.  
The rather good fitting in all the regimes shows that the local 
active force induces the polymer to behave as if it had a certain degree of 
stiffness. 
In the limit of small $\tau$, Eq.~\eqref{eq:stiff} recovers the 
Gaussian-like behavior that is consistent with the globular shape of the 
polymer (Fig.~\ref{fig:multiplot2}~(a)), while reproduces the shape of $\mathcal{P}(\mathcal{R})$ at large 
$\tau$ (Fig.~\ref{fig:multiplot2}~(b)).

\begin{figure}[!t]
\centering
\includegraphics[clip=true,width=0.8\columnwidth,keepaspectratio]
{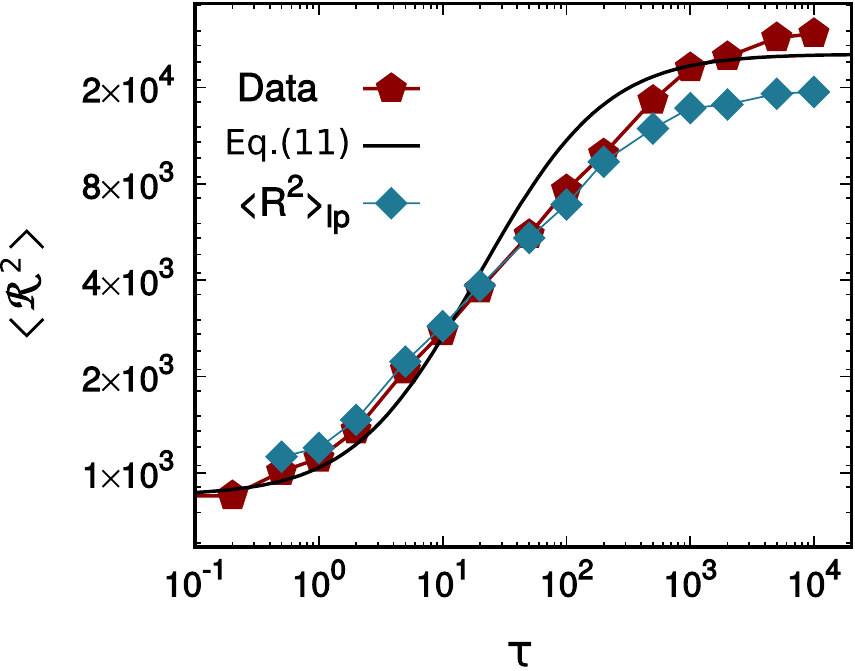}
\caption{ $\langle
 \mathcal{R}^2 \rangle$ for different 
values of $\tau$ (red pentagons) compared with 
Eq.\eqref{eq:varianceEndtoend} (black line) and the prediction, Eq.\eqref{eq:variance_stff} (light blue diamonds).
The remaining parameters are $D_a/\tau=10^2$, $\sigma=5$, $k=10$, $D_t=0.1$ and $\tau_0=0.1$.
Simulations are obtained using a time-step $\sim 10^{-4}$ and each configuration is evolved, at least, for a final time $\sim10^2\tau$.
}
\label{fig:variancetime}
\end{figure}

We use the observable $\langle\mathcal{R}^2\rangle$ as an indicator of the crossover from the compact to the elongated structures, visualized in 
Fig.\ref{fig:snapshot}.
In particular, Fig~\ref{fig:variancetime} shows the monotonic increase of $\langle\mathcal{R}^2\rangle$ as a function of $\tau$.
The phenomenological theory of stiff polymers, Eq.\eqref{eq:stiff},predicts for $\langle \mathcal{R}^2\rangle$ the expression \cite{Thirumalai}:
\begin{equation}
\langle \mathcal{R}^2\rangle_{lp} = 2 \ell_p L + 2\ell_p^2 \left( e^{-L/\ell_p} -1   \right) \,.
\label{eq:variance_stff}
\end{equation}
As expected, Eq.\eqref{eq:variance_stff} fairly agrees with data, as 
Fig.~\ref{fig:variancetime} shows, where $l_p$ is obtained from 
the numerical fit of relation~\eqref{eq:stiff}.

To attempt a theoretical prediction on $\langle\mathcal{R}^2\rangle$, 
going beyond a phenomenological theory,
we make the approximation $\sigma \simeq 0$ in the 
potential~\eqref{eq:PotentialPolymer} transforming the polymer into a  
Rouse-chain~\cite{Rouse}.
The expression of $\langle\mathcal{R}^2\rangle$ obtained for the Rouse chain
using the normal mode decomposition is (see appendix~\ref{app:A}), 
\begin{equation}
\begin{aligned}
\label{eq:varianceEndtoend}
\langle
 \mathcal{R}^2 \rangle &= 
	\dfrac{3 D_t}{\tau_0 \,k}(N-1) + \\
       & \dfrac{6 D_a \tau_0^2}{N^2} \sum_{(p,q)=1}^{N-1}
        \dfrac{c(p)c(q)\,G(p)G(q)}{\gamma_p + \gamma_q}
        \bigg[
        \dfrac{1}{1 + \gamma_p\tau_a}
        \bigg] \,,
\end{aligned}
\end{equation}
where $c(p)$ and $G(p)$ are dimensionless coefficients depending only on 
the index $p$ and on $N$: 
\begin{flalign}
G(p) &= -4 \sin\bigg(\frac{p\pi}{2}\bigg) 
	  \sin\bigg[\frac{p\pi}{2N}(N-1)\bigg]\\
c(p) &= \cos\bigg[\frac{p\pi}{2N}(2N-1)\bigg] \,.
\label{eq:c(p)}
\end{flalign}
and  $\gamma_p = 1/\tau_p$ (Eq.~\eqref{eq:taup}) has the dimension of an 
inverse time.
Eq.\eqref{eq:varianceEndtoend} contains two contributions,    
the first one, entirely due to the thermal agitation of the solvent,  
is constant and controlled by the ratio $D_t/k\tau_0$. 
The second one is due to the active force and is controlled by the ratio 
$D_a \tau_0/k/(1+ k \tau \tau_0)$.
It is straightforward to see that in the limit $\tau\to 0$ the term $D_a \tau_0 \propto \tau \tau_0$ plays the role of an effective temperature, 
in agreement with our previous discussion. 
Moreover, in the equilibrium regime, $\tau=0$, this term vanishes thus 
the well-known equilibrium result is recovered.
In particular, being the ratio $D_a/\tau$ fixed, the small-$\tau$ limit 
implies that the active force gives only a contribution of order 
$O(\tau)$ to $\langle \mathcal{R}^2 \rangle$.
We remark that the active term in Eq.~\eqref{eq:varianceEndtoend} shows a 
non-trivial dependence on each mode.
In Fig.~\ref{fig:variancetime}, we compare the $\langle\mathcal{R}^2\rangle$ 
from simulations (dots) computed at different values of $\tau$ with 
the prediction~\eqref{eq:varianceEndtoend} rescaled by the factor $\sigma^2$
since the Rouse chain turns to be more compact than the 
model~\eqref{eq:PotentialPolymer}.   
Despite the approximation, the prediction fairly 
agrees with data, both for small and large values of $\tau$.

\subsection{How the active force affects the relaxation times of the modes}
The Rouse-chain approximation allows us to study analytically 
the influence of the local active force on the chain relaxation. 
Indeed, the time correlation of the generic Rouse-mode, $C_{pp}(t,s)$, can be computed   
explicitly and reads
\begin{equation}
\begin{aligned}
C_{pp}(t,s) &= f_{\mathrm{Th}}\dfrac{e^{-\gamma_p|t-s|}}{2\gamma_p} \\
&+ \dfrac{3 D_a\tau_0^2}{N^2}c^2(p)\;\dfrac{\gamma_p\tau e^{-|t-s|/\tau}-e^{-\gamma_p|t-s|}}{\gamma_p [(\gamma_p\tau)^2 - 1]} \,.
\label{eq:Cpp(t,s)}
\end{aligned}
\end{equation}
The derivation of Eq.~\eqref{eq:Cpp(t,s)} is reported in Appendix~\ref{app:A}.
The first term represents the passive contribution of thermal agitation 
in the absence of any active source of motion.
In that case, the $p$-mode relaxation time is simply $1/\gamma_p$.

The active force gives rise to the second term in Eq.~\eqref{eq:Cpp(t,s)}, 
which is the sum of two exponentials.
The second exponential survives even in the equilibrium limit $\tau \to 0$ 
but trivially determines just a renormalization of the auto-correlation amplitude without affecting the correlation time.
Instead, in the limit of $\tau \gg 1/\gamma_p$, very interesting consequences 
emerge as the relaxation is dominated by
\begin{equation}
C_{pp}(t,s) \sim \dfrac{3 \tau_0^2 D_a c^2(p)}{N^2}\dfrac{\tau e^{-|t-s|/\tau}}{ [(\gamma_p\tau)^2 - 1]} \,,
\end{equation}
therefore, all the modes decay in the same manner.

In elongated conformations $\tau \gg 1/\gamma_p$ for every $p$, 
meaning that $\tau$ is the only relevant time in the polymer dynamics. 
Instead, in the full or partial coiled conformations,  
we have $1/\gamma_p > \tau$, at least for the lowest $p$, and the active 
force is able to affect only the dynamics of the faster modes. 

The analysis of these sections suggests that even a local active force 
on the terminal monomer is able to determine important consequences on 
the dynamics of the entire polymer, making possible drastic global 
rearrangements of its conformations.

In the next Section, we investigate the role of the active 
force at a single monomer level, finding even strong local distortions 
in the inter-monomer distances.

\subsection{Local effects of the self-propulsion on polymer structures}

\begin{figure}[!t]
\includegraphics[clip=true,width=0.8\columnwidth,keepaspectratio]
{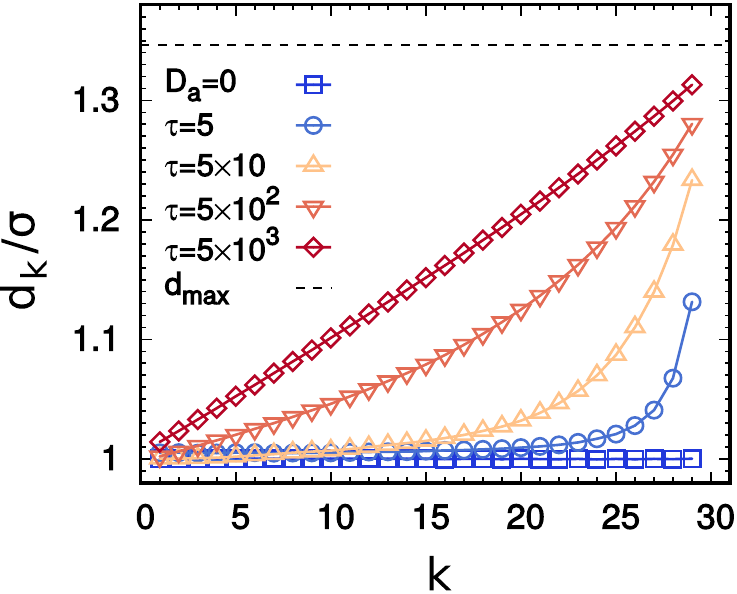}
\caption{Average distance between neighboring monomers, 
$r_k/\sigma=\langle |\textbf{r}_{k+1} - \textbf{r}_k | \rangle/\sigma$, as a function of 
the monomer index $k$ for different values of $\tau$, as shown in the legend. 
The dot black line represents the theoretical prediction for the distance 
between the head and $N-1$-th monomer, i.e. $d_{max}/\sigma$.
The remaining parameters are $D_a/\tau=10^2$, $\sigma=5$, $k=10$, $D_t=0.1$ and $\tau_0=0.1$.
Simulations are obtained using a time-step $\sim 10^{-4}$ and each configuration is evolved, at least, for a final time $\sim10^2\tau$.
}
\label{fig:dista_media}
\end{figure}

\begin{figure*}[!t]
\includegraphics[clip=true,width=0.8\textwidth,keepaspectratio]
{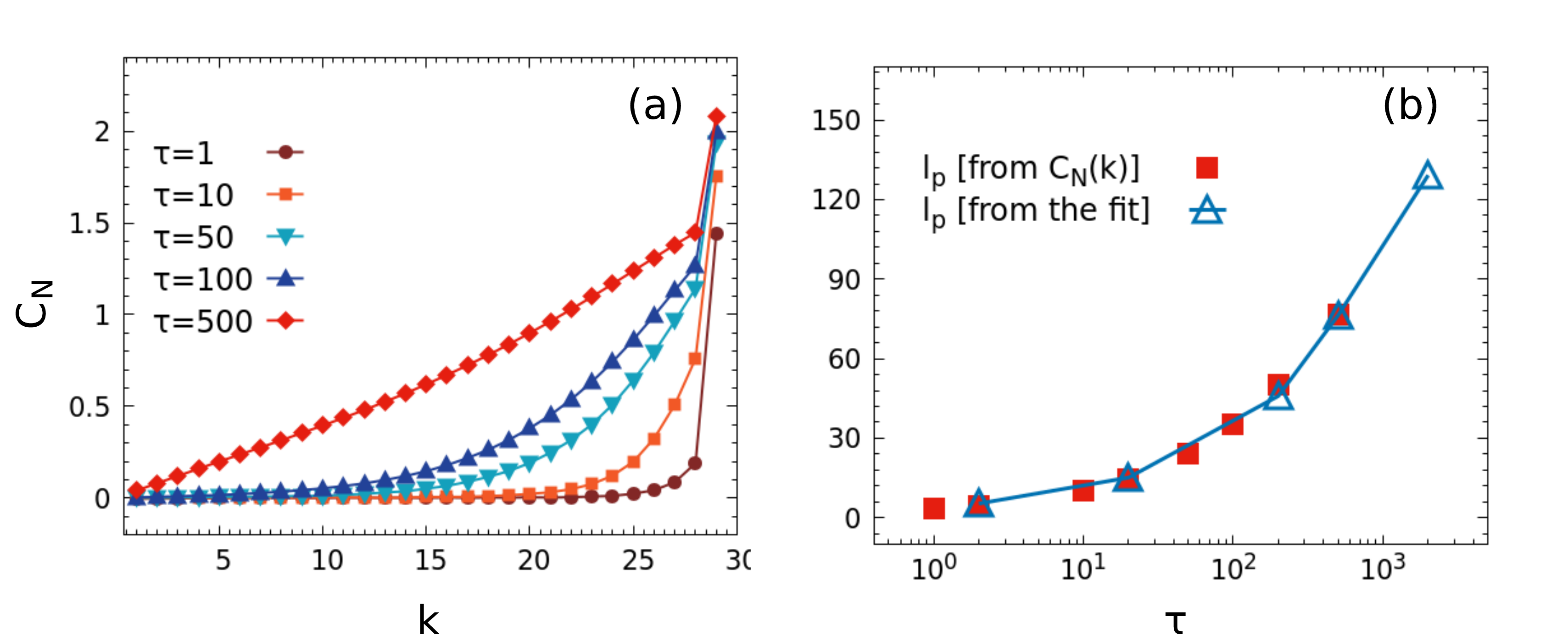}
\caption{ Panel (a): Bond-bond spatial correlation along the contour length of the polymer, $C_N(k)$ (defined by Eq.~\eqref{eq:bond-bond_corr}),  as a function of $k$, shown for different values of $\tau$. 
Panel (b): $l_p$ as a function of $\tau$, calculated from the fit (blue triangles) of the stiff polymer (Eq.~\eqref{eq:stiff}) and the best exponential fit of $C_N(k)$, as explained in the text.
The remaining parameters are $D_a/\tau=10^2$, $\sigma=5$, $k=10$, $D_t=0.1$ and $\tau_0=0.1$.
 Simulations are obtained using a time-step $\sim 10^{-4}$ and each configuration is evolved, at least, for a final time $\sim10^2\tau$.
}
\label{fig:C_N(k)}
\end{figure*}

To understand how the deformation 
induced by the active force propagates along the chain, for different 
values of $\tau$, we plot in Fig.~\ref{fig:dista_media}~(a) the average 
distance between consecutive monomers, 
$d_k=\langle |\mathbf{r}_{k+1} - \mathbf{r}_k | \rangle$, as a 
function of the monomer site $k$. 

In the regime of small active-force persistence, we find 
$d_k \simeq \sigma$, in analogy with passive polymers in solution;  
the bond fluctuations are weakly affected by the active force. 
When $\tau$ increases, the average distance between consecutive 
monomers is no longer constant because there is a transmission of the 
active-force from the head backward to the tail. Therefore, the bonds 
near the head, $N$, result stretched, $d_k \geq \sigma$,  
and the stretching degree decays to $\sigma$ for the farther monomers.

A further increase of $\tau$ is responsible for a larger stretching of the 
chain till to reach an almost linear profile 
$$
d_k \simeq \sigma + b k
$$
for $\tau \simeq 5\times 10^3$.
Large active forces on the head overstretch the polymer and induce 
a bond deformation which increases approaching the head.  
In particular, we note that the average distance between the head and the 
first passive monomer can be estimated by 
$d_{\mathrm{max}}=\sigma + \sqrt{3D_a}{\tau}/k$, thus 
$b \simeq d_{\mathrm{max}}/(N-1)$.
Such a distance can be roughly obtained by replacing $f_a$ 
with its standard deviation, an assumption which is 
meaningful as long as $\tau$ is large. 

In conclusion, we can say that the forcing effects of $\mathbf{f}_a$ 
propagates backward from the monomer $N$ along the chain establishing a 
gradient of bond deformation.
We remark that this picture is qualitatively reproduced even 
in the Rouse-chain approximation, as explicitly shown in 
Appendix~\ref{app:A}.

It is interesting to observe that the conformation of 
Fig.\ref{fig:snapshot}(b) 
suggests a phenomenology similar to the trumpet formation in polymers 
pulled by a constant force \cite{rowghanian2011force,rowghanian2012propagation} 
which is characterized by a scaling law in the tension propagation. 
The analogy with the trumpet regime is however difficult to establish on a 
quantitative basis, for two reasons: the moderate size of our chains does 
not allow this scaling to be verified, our force is not systematic and 
vanishing on the average.

We, also study the bond-bond spatial correlation along the contour length 
of the chain, referred to the terminal monomer $N$, defined as 
\begin{equation}
	C_N(k) = \langle \left({\mathbf r}_{N}-{\mathbf r}_{N-1} \right) \cdot \left({\mathbf r}_{k+1}-{\mathbf r}_{k} \right) \rangle\,.
\label{eq:bond-bond_corr}
\end{equation}
where the average is computed over stationary chain conformations. 
Figure~\ref{fig:C_N(k)}~(a) plots $C(k)$ vs $k$, for different values of 
$\tau$, revealing a monotonic increase moving towards the terminal $N$.
The growth of $C(k)$ is roughly exponential, with a typical length 
increasing with $\tau$.
To confirm the qualitative scenario of stiff polymers, we compare the 
value of $l_p$ extracted from the fit of 
$\mathcal{P}\left(\mathcal{R}\right)$ (Eq.~\eqref{eq:stiff}) 
with the correlation length associated to $C_N(k)$ and extracted from the 
best exponential fit of each curve in Fig.~\ref{fig:C_N(k)} (a).
The plot in Fig.~\ref{fig:C_N(k)} (b) shows the consistency 
of the two observables, both growing monotonically with $\tau$.
This agreement verifies the applicability of the stiff-polymers approach 
to our active chain.

\begin{figure}[!t]
\centering
\includegraphics[clip=true,width=1\columnwidth,keepaspectratio]{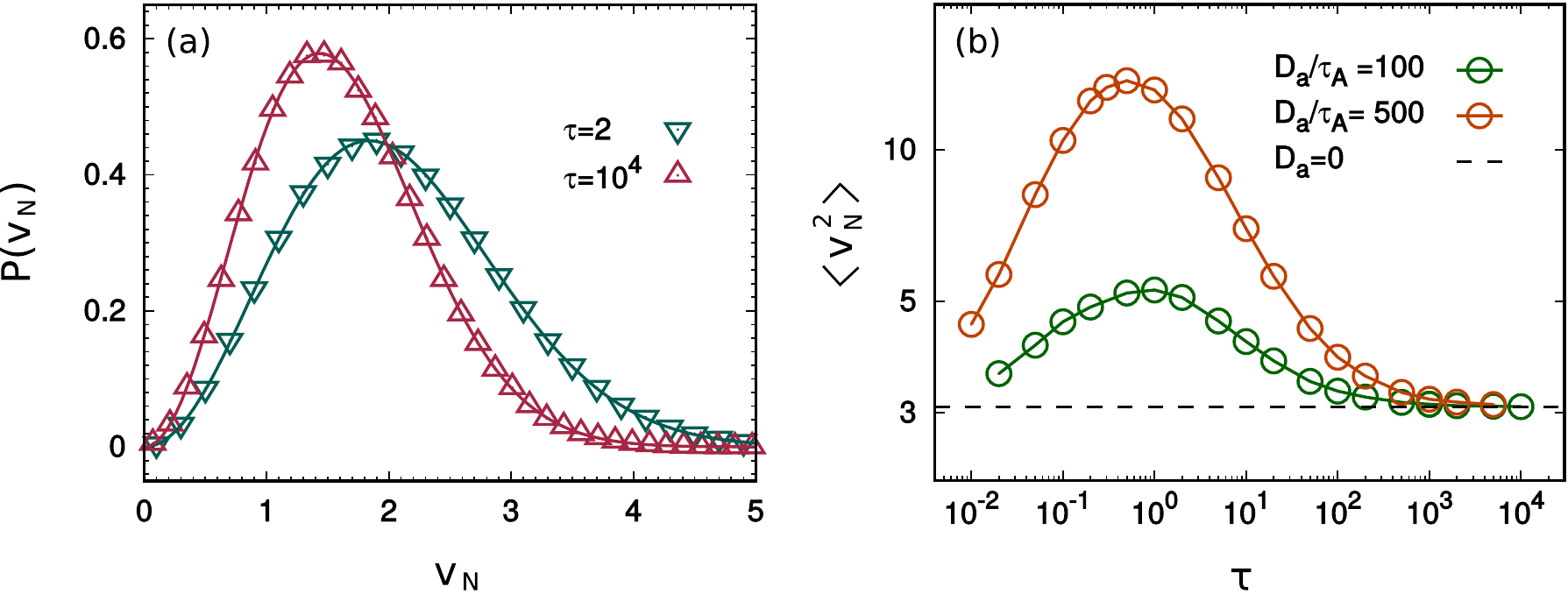}
\caption{Panel (a): Probability distribution of the head velocity, $P(|\mathbf{v}_N|)$ for two different values of $\tau$ at $D_a/\tau=10^2$.
Panel (b): variance of the distribution of $\mathbf{v}_N$ as a function of $\tau$ for two different values of $D_a/\tau$, as shown in the legend.
The remaining parameters are $\sigma=5$, $k=10$, $D_t=0.1$ and $\tau_0=0.1$.
 Simulations are obtained using a time-step $\sim 10^{-4}$ and each configuration is evolved, at least, for a final time $\sim10^2\tau$.
 }
\label{fig:vN3}
\end{figure}

In Fig.~\ref{fig:vN3}~(a), we study the modulus of the velocity probability 
distribution of the 
head, $p(|\mathbf{v}_N|)$, showing two typical shapes for a small and a 
large value of  $\tau$.
In both cases, the distributions are Gaussian-like:
\begin{equation}
P(|\mathbf{v}_N|) \propto |\mathbf{v}_N|^2 \exp{\left(- \frac{|\mathbf{v}_N|^2}{2\langle \mathbf{v}_N^2\rangle}\right)} \,,
\end{equation}
with different variances, $\langle \mathbf{v}_N^2 \rangle$, 
whose dependence on $\tau$ is reported in Fig.~\ref{fig:vN3}~(b). 
The Gaussianity is obvious in the regime of small $\tau$, in particular 
when the active force can be roughly considered as an additional 
Brownian noise.
In this regime, a first growth of $\tau$ determines an enlargement of the 
variance of the distribution, as shown in panel(b).
In particular, the variances are given by $\langle \mathbf{v}_N^2 \rangle =3 D_t/\tau_0 + 3 D_a\tau_0$. 
We note also that keeping fixed $D_a/\tau$ implies that $D_a \propto \tau$, 
which explains the initial linear growth with $\tau$ in 
Fig.~\ref{fig:vN3}~(b), until a maximal value is reached.
The expression $3 D_t/\tau_0 + 3 D_a\tau_0$ fails for $\tau \geq O(1)$ 
where the 
persistence of the motion prevents the interpretation of the active force 
as another source of diffusion.
In this regime, the variance of the distribution decreases again, until a plateau 
$\langle \mathbf{v}_N^2 \rangle = 3D_t/\tau_0$ is reached meaning that the active force does not affect the distribution of $|\mathbf{v}_N|$. 
This value of the plateau $T$ is not so surprising since $\tau \to \infty$ 
corresponds to the limit of a constant driving force, which is not expected 
to influence the fluctuations of $|\mathbf{v}_N|$. 

Such a study has revealed  a non-monotonous behavior in the variance of the 
distribution (roughly its effective temperature) which is in agreement with 
the recent observation~\cite{caprini2019activityinduced}: even for an 
interacting system of spherical particles, the increase of $\tau$ induces at 
first the warming of the system, while a further increase leads to its 
cooling.

\section{Conclusions}
In this work, we studied the transport of a Rouse-like polymer driven 
by a local active force localized in the terminal monomer (active head) 
to characterize the effects of the activity on the chain conformations. 
Upon increasing the persistence of the self-propulsion,
we observed a transition from globular to open conformations,
revealing the presence of a regime where random-coil and partially
elongated conformations coexist.
This transition is well-described by the statistics of the end-to-end distance, in particular its distribution and the second moment, 
whose numerical analysis is supported by theoretical predictions provided  
by stiff-polymer theory and Rouse-chain calculations.
Moreover, we investigated the local properties of the chain
focusing both on bond stretching and bond-bond
correlation along the contour length, in fair agreement with the 
phenomenological stiff-polymer theory.
We find that the active force acting on the head
induces a gradient of bond deformation in regimes of strong persistence,
deeply affecting the ``microscopic'' structural properties of the chain.
Our results could be easily generalized to the case of more complex potentials going beyond the simple linearity of the Rouse-chain. Different shapes of the potential lead to the same phenomenology even if the bond stretching could be consistently reduced choosing a stiffer attraction between monomers.

Our study is a contribution towards the comprehension of the complex 
interplay between shape-deformability and local self-propulsion in extended
systems.
Generalizing this study to more complex and deformable geometries could 
represent a very promising point to go beyond the approximation of 
self-propelled rigid bodies.

\section*{Acknowledgment}
This research was supported by M.I.U.R., Prin  
``Coarse-grained description for non-equilibrium systems and 
transport phenomena'' (CO-NEST), Grant n.201798CZLJ.

\section*{Conflicts of interest}
There are no conflicts of interest to declare.


\begin{widetext}

\appendix
\section{Strength and persistence time of $\mathbf{f}_a$}

In this paper, we employ the AOUP model to reproduce the self-propulsion
in three dimensions.
This is modeled as a time-dependent force evolving by three independent O.U. processes, i.e. by Eq.\eqref{eq:activeforcedynamics}.

As already explained in the text, $\tau$ is the persistence time of 
the dynamics determining also the correlation time of the auto-correlation 
of $\mathbf{f}_a$, which reads:
$$
\left\langle \mathbf{f}_a(t) \cdot \mathbf{f}_a(s) \right\rangle = 3\frac{D_a}{\tau} \exp{\left(- \frac{\left| t-s \right|}{\tau}   \right)} \,.
$$
Additionally, since the steady state solution of Eq.\eqref{eq:activeforcedynamics} (i.e. the marginal probability distribution of $\mathbf{f}_a$) is:
$$
p\left(\mathbf{f}_a\right) \propto \exp{\left( - \frac{D_a}{\tau}\frac{\left|\mathbf{f}_a\right|^2}{6}    \right)} \,,
$$
it is straightforward to conclude that the square root variance of the active force, which is proportional to $D_a/\tau$, determines the strength of the self-propulsion force, being $\langle\left| \mathbf{f}_a \right| \rangle = \sqrt{D_a/\tau}$.

\section{Velocity distribution of the center of mass\label{app:velocity}}
The dynamics of the center of mass, Eqs.~\eqref{eq:centerofmass_maintext}, 
and of the self-propulsion, Eqs.~\eqref{eq:centerofmass_maintext}, can be analyzed by deriving the associated Fokker-Planck equation 
\cite{Gardiner} governing the evolution of the probability distribution 
$f(\mathbf{x}_c, \mathbf{v}_c, \mathbf{f}_a, t)$, 
\begin{equation}
\begin{aligned}
\partial_t f &=  \frac{\nabla_{\mathbf{v}_c}}{\tau_0}\cdot \left( \mathbf{v}_c f - \frac{\tau_0}{N} \mathbf{f}_a f\right) + \frac{D_t}{N\tau_0^2} \nabla^2_{\mathbf{v}_c} f  -\mathbf{v}_c \cdot \nabla_{\mathbf{x}_c} f+\nabla_{\mathbf{f}_a}\cdot\left(\frac{\mathbf{f}_a}{\tau}f\right) +\frac{D_a}{\tau^2}\nabla^2_{\mathbf{f}_a}f
\end{aligned}
\end{equation}
where $\nabla$ and $\nabla^2$ indicate the gradient and the Laplacian operator, with respect to the variables in the subscript. 
These equation is diffusive in space, but admits a steady-state
distribution, $p(\mathbf{v}_c, \mathbf{f}_a)$, in velocity and self-propulsion. 
The linearity of the process implies that $p(\mathbf{v}_c, \mathbf{f}_a)$ is a multivariate Gaussian 
\begin{equation*}
\begin{aligned}
&p(\mathbf{v}_c, \mathbf{f}_a) \propto G(\mathbf{f}_a)\;\exp{\left[-\frac{\beta_{\mathrm{eff}}}{2}\left( \mathbf{v}_c - \frac{\mathcal{C}}{N}\,\mathbf{f}_a   \right)^2\right]}\,,\\
&G(\mathbf{f}_a) \propto \exp{\left[ - \left(1+\frac{\Gamma^4}{\left(1+\Delta\right)^2}\frac{\tau}{\tau_0} \dfrac{1}{\frac{1}{\frac{1}{\Gamma}+\Delta}+\frac{\tau}{\tau_0}\frac{\Gamma^3}{(1+\Delta)^2}}\right)\frac{\tau}{D_a}\frac{\mathbf{f}_a^2}{2\,N}  \right]} \,,\\
\end{aligned}
\end{equation*}
where the coefficients reads:
\begin{flalign*}
&\beta_{eff} = \frac{1}{\tau_0 D_a}\left(\dfrac{1}{\frac{1}{\Gamma}+\Delta}+\frac{\tau}{\tau_0} \frac{\Gamma^3}{\left( 1+\Delta\right)^2}   \right) \,,\\
&\mathcal{C}= \frac{\Gamma^2}{1+\Delta}  \dfrac{1}{\dfrac{1}{\frac{1}{\Gamma}+\Delta}+\frac{\tau}{\tau_0} \frac{\Gamma^3}{\left( 1+\Delta\right)^2}   } \,,
\end{flalign*}
being $\Gamma=1+\tau/\tau_0$ and $\Delta=D_tN /(\tau_0^2 D_a)$.

\section{Rouse-mode analysis of correlations\label{app:A}}
Rouse model, whose potential energy is Eq.\eqref{eq:PotentialPolymer} with 
$\sigma = 0$, can be analytically solved by a decompositions in Rouse-modes
\begin{equation}
{\mathbf{r}_n(t)} = {\mathbf X_0(t)} +  2\sum_{p=1}^{N-1} 
{\mathbf X_p(t)} \cos\bigg[\frac{p\pi}{N}\bigg(n-\frac{1}{2}\bigg)\bigg]
\label{eq:app_rX}
\end{equation}
where each mode, defined as   
$$
{\mathbf X}_p(t) = \frac{1}{N} \sum_{n=1}^{N} {\mathbf r}_n(t) 
\cos\bigg[\frac{p\pi}{N}\bigg(n-\frac{1}{2}\bigg)\bigg]\,,
$$
is independent of the others and evolves, in the stationary regime, 
according to 
\begin{equation}
{\mathbf X_p(t)} = \tau_0 e^{-\gamma_p t}
\int_{-\infty}^{t}\!\!ds\; e^{\gamma_p s}
[{\mathbf F_p(s)} + {\mathbf A_p(s)}] \,,
\label{eq:modes}
\end{equation}
with ${\mathbf F_p}$ and ${\mathbf A_p}$ the mode components of the Brownian 
and active force, respectively. Instead, $\gamma_p$ is given by
$$
\gamma_p = 4 \,k\, \tau_0   
\sin^2 \bigg(\frac{p\pi}{2N}\bigg)\,,
$$
being $k$ the stiffness of the chain and $\tau_0$ the relaxation time 
of the solvent. 

A central quantity of our approach is the stationary time-correlation 
between the modes for $p>0$ and $q>0$, i.e.
$$
C_{pq}(t-s) = \langle {\mathbf X}_p(t) {\mathbf X}_q(s) \rangle  \,,
$$
that can be evaluated by using Eq.\eqref{eq:modes}, 
\begin{equation*}
\begin{aligned}
&\langle {\mathbf X_p(t)}\cdot{\mathbf X_q(s)} \rangle =
 \tau^2_0  e^{-(\gamma_p t  +\gamma_q)s}
\int_{-\infty}^{t}\!\!du\int_{-\infty}^{s}\!\!dv
	e^{\gamma_p u + \gamma_q v'}\bigg[
      \langle {\mathbf F_p(u)}\cdot{\mathbf F_q(v)}\rangle + 
	      \langle {\mathbf A_p(u)}\cdot{\mathbf A_q(v)}\rangle
\bigg] \,.
\end{aligned}
\end{equation*}
The correlations of the Fourier components of the thermal and active noises
can be easily derived from the direct correlation of such noises considered 
in Eqs.\eqref{eq:polymerdynamics},
\begin{flalign}
&\langle {\mathbf F_p(s)}\cdot{\mathbf F_q(s')}\rangle = 
	\label{corr1}
\dfrac{3D_t}{N\tau_0^2} \bigg[ \delta(p+q) + \delta(p-q) \bigg]\delta(s'-s)  \\
	\label{corr2}
&\langle {\mathbf A_p(s)}\cdot{\mathbf A_q(s')}\rangle = 
\dfrac{3 D_a}{N^2\tau} c(p) c(q) \exp(-|s'-s|/\tau) \,.
\end{flalign}
Thus $C_{pq}(t-s)$ is expressed by the sum of two terms 
$$
C_{pq}(t-s) = f_{\mathrm{Th}}\delta_{p,q}\dfrac{e^{-\gamma_p|t-s|}}{2\gamma_p}
+ f_{\mathrm{Act}} c(p) c(q) S_{pq}(t-s); 
$$
where  
$$
c(p)= \cos\bigg[\frac{p\pi}{2N}(2N-1)\bigg] = 
(-1)^p \cos\bigg[\frac{p\pi}{2N}\bigg] \,.
$$
and the two constants are defined as
$$
f_{\mathrm{Th}} = \dfrac{3 D_t}{N}  \quad,\quad 
f_{\mathrm{Act}} = \dfrac{3 D_a \tau_0^2}{N^2}  \,.
$$
The last term in $C_{pq}(t-s)$ refers to the active force and 
contains the integral 
$$
S_{pq}(t-s) = \frac{1}{\tau}
\int_{-\infty}^{t}\!\!du\int_{-\infty}^{u}\!\!du 
e^{\gamma_p(u-t) + \gamma_q(v-s)}
e^{-|u-v|/\tau}
$$
which can be solved providing the result 
\begin{equation}
S_{pq}(t-s) =
\begin{cases} 
\dfrac{\tau e^{-|t-s|/\tau}}{(\gamma_q\tau +1)(\gamma_p\tau -1)} - 
\dfrac{2 e^{-\gamma_p|t-s|}}{(\gamma_p + \gamma_q)(\gamma_p^2\tau^2 -1)} & 
        t>s\\
        & \\
        \dfrac{\tau e^{-|t-s|/\tau}}{(\gamma_p\tau +1)(\gamma_q\tau - 1)} -
\dfrac{2 e^{-\gamma_q|t-s|}}{(\gamma_p + \gamma_q)(\gamma_q^2\tau^2 -1)} &
        t<s 
\end{cases}.
\label{eq:Spq}
\end{equation}
When $p=q>0$, we are left with the mode-mode autocorrelation which simplifies 
to
$$
C_{pp}(t,s) = f_{\mathrm{Th}}\dfrac{e^{-\gamma_p|t-s|}}{2\gamma_p} + 
f_{\mathrm{Act}} c^2(p)\;\dfrac{\gamma_p\tau e^{-|t-s|/\tau}-e^{-\gamma_p|t-s|}}{\gamma_p [(\gamma_p\tau)^2 - 1]} \,.
$$
In the following, we will need the correlation at the same time, $t=s$,
a quantity that, in the stationary regime, becomes independent of time and reads 
\begin{equation}
C_{pq}(0) = f_{\mathrm{Th}}\dfrac{\delta_{p,q}}{2\gamma_p} +  
	f_{\mathrm{Act}} c(p)c(q)\; S_{pq}(0),
\label{eq:Corr}
\end{equation}	
with the obvious notation [see. Eq.\eqref{eq:Spq}]
$$
S_{pq}(0) = \dfrac{1}{\gamma_p + \gamma_q} 
\bigg[\dfrac{1}{\gamma_p\tau+1} + \dfrac{1}{\gamma_q\tau+1}\bigg]\,.
$$

\subsection{Center of mass behaviour}

\begin{figure*}[!t]
\includegraphics[clip=true,width=0.8\textwidth,keepaspectratio]
{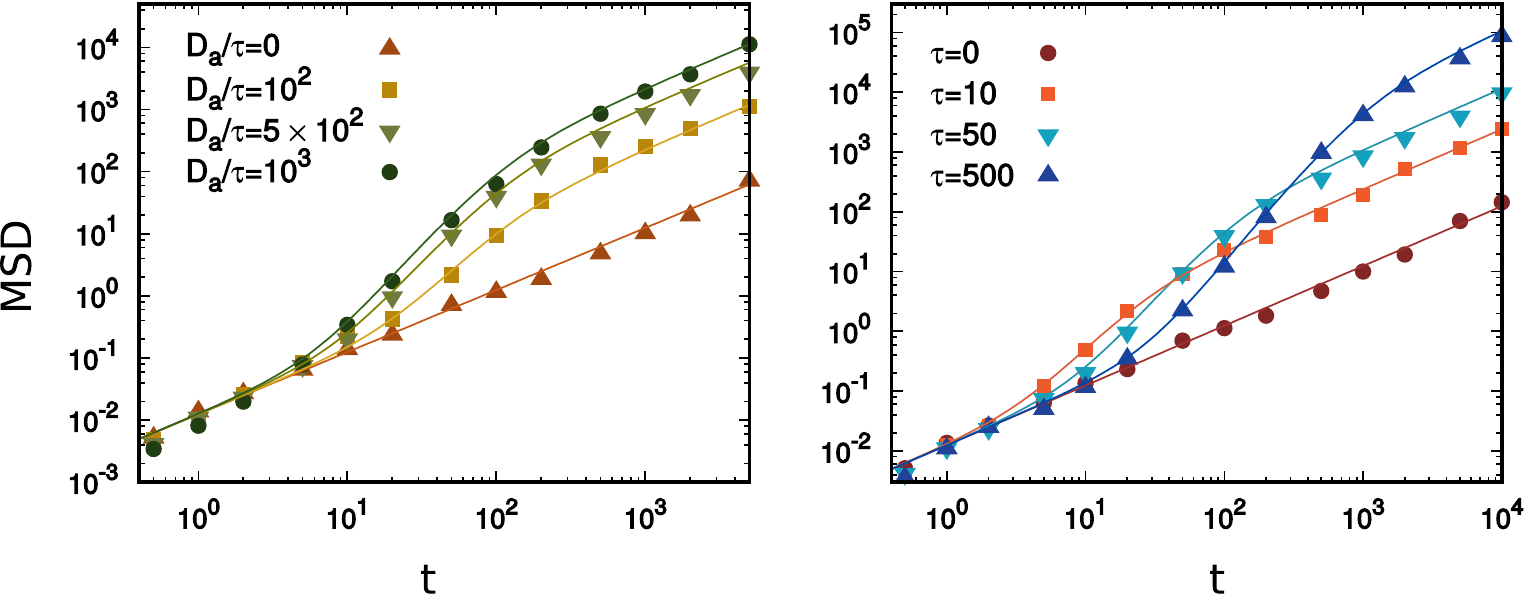}
\caption{
$MSD(t)$ for different values of the active force parameters.
In panel (a) we adopt the protocol i), namely varying $D_a/\tau$ at fixed $\tau=50$, while we employ the protocol ii) in panel (b), i.e. varying $\tau$ keeping fixed $D_a/\tau=500$. 
The remaining parameters are $\sigma=5$, $k=10$, $D_t=0.1$ and $\tau_0=0.1$.
 Simulations are obtained using a time-step $\sim 10^{-4}$ and each configuration is evolved, at least, for a final time $\sim10^2\tau$.
}
\label{fig:MSD}
\end{figure*}

The center of mass, corresponding to the mode $p=0$, and given by
$$
{\mathbf X_0(t)} = \frac{1}{N} \sum_{n=1}^{N} {\mathbf{r}_n(t)}
$$
evolves in time from the initial condition ${\mathbf X_0(0)}$ as 
\begin{equation}
{\mathbf X_0(t)} = {\mathbf X_0(0)} + 
\tau_0\int_{0}^{t}\!\!ds\; 
[{\mathbf F_p(s)} + {\mathbf A_p(s)}] \,.
\label{eq:app_Rousemodessol}
\end{equation}
Therefore, its mean square displacement (MSD) will behave as
\begin{align} 
\label{eq:app_MSD}
	 \langle [{\mathbf X_0(t)} - {\mathbf X_0(0)}]^2 \rangle = 
	 \tau^2_0 \int_{0}^{t}\!\!ds \int_{0}^{t}\!\!ds'\; 
\bigg[
   \langle{\mathbf F_0(s)}\cdot{\mathbf F_0(s')}\rangle +   
   \langle{\mathbf A_0(s)}\cdot{\mathbf A_0(s')}\rangle     
\bigg]  \nonumber\,.
\end{align}
The two-time correlations involved are given by Eqs.(\ref{corr1},\ref{corr2}) 
taken for $p=q=0$
\begin{eqnarray*}
  \langle{\mathbf F_0(s)}\cdot{\mathbf F_0(s')}\rangle &=& 
  \frac{6 \, D_t}{N \tau_0^2}\delta(s-s') \\
  \langle{\mathbf A_0(s)}\cdot{\mathbf A_0(s')}\rangle &=&
	\frac{3 D_a}{N^2 \tau}\exp\bigg(-\frac{|s-s'|}{\tau}\bigg) \,,
\end{eqnarray*}
Thus, Eq.\eqref{eq:app_MSD} turns into
$$
{\mathrm{MSD}}(t) =  
\frac{6 D_t}{N} t +  \tau^2_0 \frac{3 D_a}{N^2\tau} 
\int_{0}^{t}\!\!ds \int_{0}^{t}\!\!ds'\; e^{-|s-s'|/\tau}    \,.
$$
In particular,  performing the integrals, we get the time-dependent expression for the MSD: 
$$
{\mathrm{MSD}}(t) =  6 \bigg(\frac{D_t}{N} +  
\frac{D_a\tau^2_0}{N^2}\bigg)\;t + 
6 \frac{D_a \tau^2_0}{N^2} \tau \big(e^{-t/\tau} - 1\big) \,.
$$
Expanding in power of $t/\tau$ we can estimate the relevance of the active force for small $t$. In particular, we get
\begin{equation*}
{\mathrm{MSD}}(t) \approx 6 \frac{D_t }{N} t + 
3 \frac{D_a \tau^2_0}{N^2} \frac{t^2}{\tau} \,.
\end{equation*}
Comparing the amplitudes of the two terms, we have a necessary condition 
to establish the relevance of the active force to the
center of mass motion of the polymer even at early stages. 

The exact expression for the $MSD(t)$, 
Eq.\eqref{eq:MSDmaintext}, is supported by a numerical comparison 
via two protocols: i) constant $\tau$ varying the ratio $D_a/\tau$, 
ii) varying $\tau$ keeping fixed $D_a/\tau$.
In both cases, the active force increases monotonically the diffusivity 
and the agreement between the two sets of data is fairly good as revealed 
by Fig.\ref{fig:MSD} (a) and (b), respectively.

\subsection{End-to-End Distance}
In the active Rouse-model approximation, we can compute analytically 
the end-to-end distance, 
$\boldsymbol{\mathcal{R}}(t) = {\mathbf{r}_N}-{\mathbf{r}_1}$,
of the Rouse Polymer, which reads:
\begin{eqnarray*}
\boldsymbol{\mathcal{R}}(t) =  
2\sum_{p=1}^{N-1} {\mathbf X_p(t)} 
\bigg(\cos\bigg[\frac{p\pi}{2N}(2N-1)\bigg] - 
      \cos\bigg[\frac{p\pi}{2N}\bigg]\bigg) \,.
\end{eqnarray*}
This equation can be written formally as the series
$$
\boldsymbol{\mathcal{R}}(t) = \sum_{p=1}^{N-1} G(p) {\mathbf X_p(t)} \,,
$$
where 
$$
G(p) =  [(-1)^p -1] \cos\bigg(\frac{p\pi}{2N}\bigg)\,. 
$$
Therefore, its variance can be expressed in terms of the stationary 
correlation of the modes at the same time 
\begin{equation}
\langle \mathcal{R}^2(t)\rangle = \sum_{(p,q)=1}^{N-1} G(p) G(q) 
\;\langle {\mathbf X_p(t)}\cdot{\mathbf X_q(t)} \rangle  \,,
\label{eq:Ree2}
\end{equation}
where $\langle {\mathbf X_p(t)}\cdot{\mathbf X_q(t)} \rangle$ 
is nothing but Eq.\eqref{eq:Corr}. 
Thus we obtain the mean square end-to-end distance, i.e. the second moment of $\mathcal{P}(\mathcal{R})$:
\begin{equation*}
\begin{aligned}
	\langle
	{\mathcal R}^2(\infty) \rangle = f_{\mathrm{Th}} \sum_{p=1}^{N-1}
        \frac{G^2(p)}{2\gamma_p}  + f_{\mathrm{Act}} \sum_{(p,q)=1}^{N-1} 
	\dfrac{c(p)c(q)\,G(p)G(q)}{\gamma_p + \gamma_q} 
	\bigg[ 
	\dfrac{1}{1 + \gamma_p\tau}+ 
	\dfrac{1}{1 + \gamma_q\tau}
	\bigg]  \,.
\end{aligned}
\end{equation*}
The symmetry in $p,q$ implies
that the expression can be recast into
\begin{equation*}
\begin{aligned}
        \langle
        {\mathcal R}^2 \rangle &= f_{\mathrm{Th}} \sum_{p=1}^{N-1}
        \frac{G^2(p)}{2\gamma_p}  
        + 2 f_{\mathrm{Act}} \sum_{(p,q)=1}^{N-1}
        \dfrac{c(p)c(q)\,G(p)G(q)}{\gamma_p + \gamma_q}
        \bigg[
        \dfrac{1}{1 + \gamma_p\tau_a}
        \bigg] \,.
\end{aligned}
\end{equation*}
It can be shown that for a Rouse-Chain the following sum-rules 
hold true
\begin{flalign*}
\sum_{p=1}^{N-1} \frac{G^2(p)}{2\gamma_p} &= 
	\dfrac{N(N-1)}{\tau_0 k} \,\\
	\sum_{p=1}^{N-1} G^2(p) &= 4N \,.
\end{flalign*}
Thus, the final expression for the mean square end-to-end distance 
reads
\begin{equation*}
\begin{aligned}
\langle
 {\mathcal R}^2 \rangle = 
	\dfrac{3 D_t}{k}(N-1) + 
        \dfrac{6 D_a \tau_0^2}{N^2} \sum_{(p,q)=1}^{N-1}
        \dfrac{c(p)c(q)\,G(p)G(q)}{\gamma_p + \gamma_q}
        \bigg[
        \dfrac{1}{1 + \gamma_p\tau}
        \bigg] \,,
\end{aligned}
\end{equation*}
which coincides with the result of the main text.
  

\subsection{Fluctuation of the bond deformation}
With the same strategy applied to derive the End-to-End distance, 
we can compute the fluctuation of the bond deformation,  
$\delta {\mathbf r}_n = {\mathbf r}_{n+1}-{\mathbf r}_n$,
induced by the active force along the Rouse chain, i.e. 
$$
\delta {\mathbf r}_n = 
2\sum_{p=1}^{N-1} {\mathbf X_p(t)} 
\cos\bigg[\frac{p\pi}{N}\bigg(n+\frac{1}{2}\bigg)\bigg] - \cos\bigg[\frac{p\pi}{N}\bigg(n-\frac{1}{2}\bigg)\bigg] \,.
$$
In a more explicit form, it can be expressed as
$$
\delta {\mathbf r}_n = -4 \sum_{p=1}^{N-1}{\mathbf X_p(t)} \sin\bigg(\frac{p\pi}{2N}\bigg) \sin\bigg(\frac{p\pi}{N}n\bigg)  \,.
$$
For the sake of shortness, it is convenient to set $G_n(p) = 4\sin(p\pi/2N)\sin(q\pi n/N)$.
Squaring and averaging, we obtain
$$
\langle (\delta{\mathbf r}_{n})^2 \rangle = \sum_{pq} \langle {\mathbf X}_p(t) {\mathbf X}_q(t) \rangle G_n(p)G_n(q) \,.
$$
Since the bond fluctuation depends on the correlation at the same time, it can be rewritten in terms of Eq.\eqref{eq:Corr}
$$
\langle (\delta{\mathbf r}_{n})^2 \rangle = 
f_{\mathrm{Th}} \sum_{p=1}^{N-1}\dfrac{G_n(p)}{2\gamma_p}
+
f_{\mathrm{Act}}\sum_{(p,q)=1}^{N-1} c(p)c(q)S_{pq}(0)G_n(p) G_n(q) \,.
$$
After some simple algebraic manipulations, and using the definition of $c(p)$, $G_n(p)$ and
$\gamma_p$, we obtain the long expression
\begin{equation*}
\begin{aligned}
\langle (\delta{\mathbf r}_{n})^2 \rangle = &\dfrac{3 D_t}{k\tau_0} + 
	4 f_{\mathrm{Act}} \sum_{(p,q)=1}^{N-1} (-1)^{p+q} S_{pq}(0) \sin\bigg(\frac{p\pi}{N}\bigg) \sin\bigg(\frac{q\pi}{N}\bigg) \sin\bigg(\frac{p\pi}{N}n\bigg) \sin\bigg(\frac{q\pi}{N} n\bigg) \,.
\label{eq:app_Rouse_intermononerdistance}
\end{aligned}
\end{equation*}
Finally, we can estimate $\langle |{\mathbf r}_{n+1} -{\mathbf r}_{n} |\rangle \approx 
\sqrt{\langle (\delta{\mathbf r}_{n})^2 \rangle}$.
As shown in Fig.\ref{fig:appgraph}, this expression reproduces qualitatively the behavior reported in Fig.\ref{fig:dista_media} despite the employment of the Rouse-approximation.
We remark that Eq.~\eqref{eq:app_Rouse_intermononerdistance} decays to a value smaller than $\sigma$ for monomers far from the head at variance with Fig.\ref{fig:dista_media}.
This is not surprising since, in the Rouse approximation, $\sigma$ does not play any role and, thus, even the passive polymer assumes a more compact configuration.
\begin{figure}[!t]
\includegraphics[clip=true,width=0.95\columnwidth,keepaspectratio]
{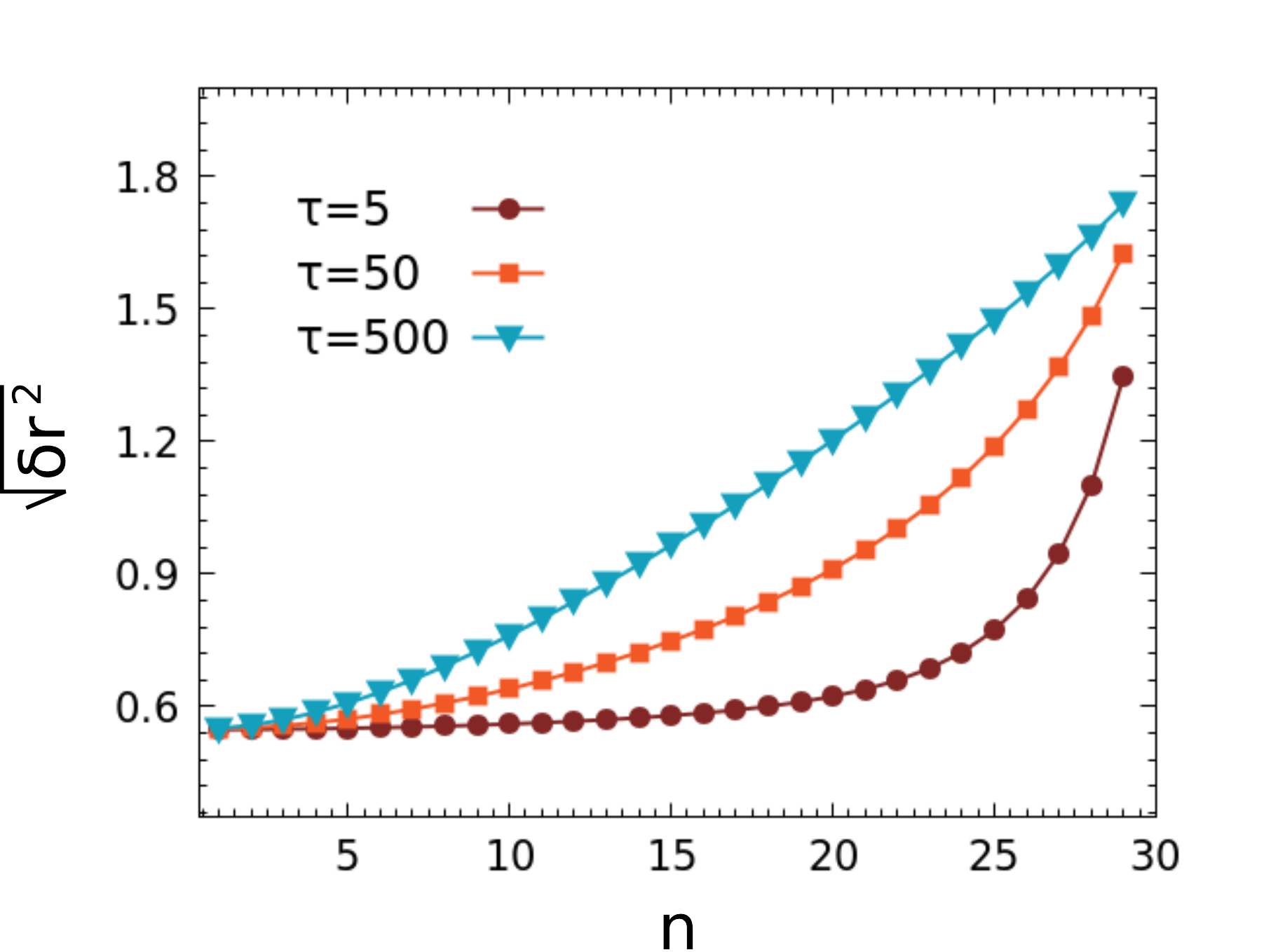}
\caption{
Eq.~\eqref{eq:app_Rouse_intermononerdistance} as a function of $n$ for three different values of $\tau$ as shown in the legend. The remaining parameters are $D_a/\tau=10^2$, $\sigma=5$, $k=10$, $D_t=0.1$ and $\tau_0=0.1$.
 Simulations are obtained using a time-step $\sim 10^{-4}$ and each configuration is evolved, at least, for a final time $\sim10^2\tau$.}
\label{fig:appgraph}
\end{figure}

\end{widetext}

\bibliographystyle{rsc} 

\bibliography{biblio} 

\end{document}